\documentclass[doublecol,amsmath,amssymb]{epl2} 
\usepackage{amsmath}
\usepackage{graphics}
\usepackage[colorlinks=true,citecolor=blue,linkcolor=red,linktocpage=true,pagebackref=false]{hyperref}

\title{Single-particle interference versus two-particle collisions}
\author{Stefan Juergens\inst{1}, 
Janine Splettstoesser\inst{1}, 
Michael Moskalets\inst{1,2}}

\institute{                    
  \inst{1} Institut f\"{u}r Theorie der Statistischen Physik, RWTH Aachen University, D-52056 Aachen, Germany and JARA - Fundamentals of Future Information Technology\\
   \inst{2} Department of Metal and Semiconductor Physics, National Technical University "Kharkiv Polytechnical Institute", 61002 Kharkiv, Ukraine
}

\abstract{
We consider a mesoscopic circuit in the quantum Hall effect regime comprising two synchronized single-particle sources emitting particles into a Mach-Zehnder interferometer. 
While particles from one source can possibly interfere at the interferometer output, particles from the second source are injected directly into one of the interferometer's arms and are used to create tunable and coherent suppression of interference.
If particles from the two different sources collide at the interferometer output the magnetic-flux dependence of the charge transferred to one of the output contacts is suppressed.
In contrast the interference pattern in the current at a fixed time is preserved and the impact of the second source manifests itself in a time-dependent phase-shift. 
}

\begin{document}

\maketitle

\section{Introduction} The particle-wave dualism lies in the heart of  Quantum Mechanics. 
In mesoscopics the discreteness of the charge leads, e.g., to the well known Coulomb blockade effect, observable in a current passing through a small metallic island or a quantum dot~\cite{Averin91}. 
The wave-like behavior of an electron is manifested in the single-particle interference leading to Aharonov-Bohm (AB) oscillations of the current through a ring~\cite{Webb85}. 
However, in these examples particle- and wave-like behavior appear in different experimental set-ups. 
The fact that electrons can show - at least partial - interference, even in the strong Coulomb blockade regime (see e.g. ref.~\cite{Konig02} and references therein) was shown in a quantum dot embedded in an Aharonov-Bohm ring~\cite{Yacoby95}. 
Our aim is different: we show that the use of the recently realized tunable single-electron sources~\cite{Feve07}, supplying particles periodically one by one, allows to observe both aspects of the quantum nature of electrons within the same electronic circuit. This is revealed when the magnetic-flux dependence of two different quantities - the transferred charge and the time-resolved current - are measured within the same interferometer setup into which single particles are injected.  
Alternative single-electron sources, likewise working in the  gigahertz regime, were experimentally realized, as shown in refs.~ \cite{Blumenthal07,Fujiwaral08,Kaestnerl08,Chan11}.

%%%%%%%%%%%%%%%%%%%%%%%%%%%%%%%%%%%%%%%%%%%%%%%%%%%
\begin{figure}[t]
\includegraphics[width=3.3in]{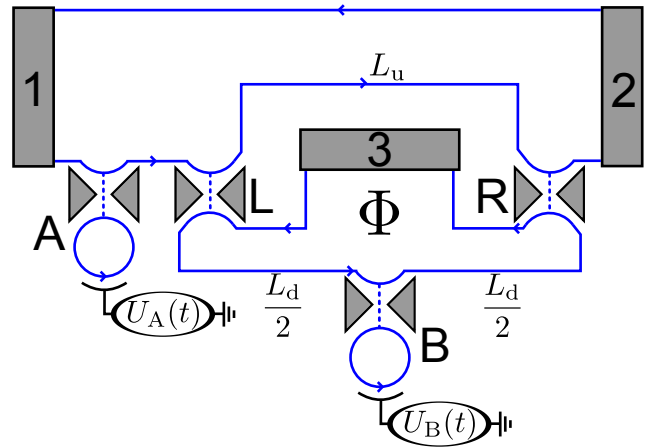}
\caption{ 
(color online). Setup of the MZI with magnetic flux $\Phi$ in the integer quantum Hall regime with unbiased reservoirs $1,2$ and $3$.  Emitters A and B are driven by  gate potentials, which are periodic in time and inject single particles into edge states (full, blue lines). QPCs L and R figure as beam splitters.
\label{fig_setup}  } 
\end{figure}
%%%%%%%%%%%%%%%%%%%%%%%%%%%%%%%%%%%%%%%%%%%%%%%%%%%

The proposed Mach-Zehnder interferometer (MZI)~\cite{Ji03},  fed by two single-particle sources, is shown in fig.~\ref{fig_setup}. 
The sources are periodically driven and we assume the regime when each source A and B is emitting one electron at a certain time during one half of the driving cycle and one hole during the other half.
On one hand, we propose to consider the time-resolved current $I_{2}(t)$ measured at the drain contact $ 2$. 
On the other hand, we contrast the time-resolved current with the charge $Q_2$, arriving at contact 2, in \textit{one half cycle}, which counts electrons arriving from the two sources at contact $2$. 
The time-resolved current, $I_{2}(t)$, and the charge $Q_{2}$ - both expectation values, meant to be measured over many cycles - show a qualitatively different behavior as a function of the magnetic flux penetrating the interferometer. 

If only source A is working, charge and current show the AB effect~\cite{Haack11}.
In this Letter, the effect of the source B on the interference pattern is of importance, where B is driven independently from source A but can be synchronized with respect to it.
The current $I_{2}(t)$ always shows the AB effect, hence indicating the wave-like behavior of electrons.
In contrast, whether the detected charge $Q_{2}$ shows the AB effect or not, depends on the occurrence of collisions between particles, emitted from A and B, at the interferometer's output, namely at the quantum point contact R (QPC R), see fig.~\ref{fig_setup}. 
The ability to collide emphasizes the particle nature of electrons.
The time of emission from the two sources is tunable and one can therefore force electrons to either collide at the QPC R or to pass it independently.
If the particles collide at the QPC R, they become Fermi-correlated and necessarily go to different contacts~\cite{Olkhovskaya08} independently of the magnetic flux. If particles from A and B do not collide, then $Q_2$ shows the AB effect due to particles stemming from source A, which remain sensitive to the magnetic flux.

We stress that no dephasing processes \cite{Chang08,Roulleau08} are involved in the interference suppression revealed here. 
The mechanism  is entirely coherent, since the time-resolved current $I_{2}(t)$ depends on the AB flux, demonstrating the absence of dephasing in the system.  

It is widely accepted that noise~\cite{Blanter00} is the minimal order current correlator, demonstrating two-particle physics 
\cite{Buttiker91,Buttiker92,Oliver99,Henny99,Samuelsson04,Beenakker06,Neder07,Splett09,Chirolli11}
 in non-interacting mesoscopic systems. 
The same is true for analogous setups in optics, see e.g.
Refs~\cite{hong87,pan08}, where two-particle collisions (of photons) are
visible in the fourth-order interference only.

Our results show that the combination of interference effects and time-dependent particle sources allows to reveal two-particle physics already with a measurement of the charge, which - for the non-interacting electrons we are looking at - is essentially a single-particle quantity.\footnote{Such a measurement could be easier compared to a noise measurement.}

%%%%%%%%%%%%%%%%%%%%%%%%%%%%%%%%%%%%%%%%%%%%%%%%%%%
\section{Model}
We study an MZI, realized in a two-dimensional electron gas in a high magnetic field - the integer quantum Hall effect regime~\cite{Klitzing80} - where transport takes place along chiral edge states~\cite{Halperin82, Buttiker88}. 
The setup is shown in fig.~\ref{fig_setup}. 
The reservoirs $1$, $2$, and $3$ are not biased. Instead, single particles, electrons and holes, are injected into the device from two single-particle emitters A and B, depicted as circular edge states in the figure. 
Particles injected from source A are reflected or transmitted at QPC L and can then travel along an upper arm with length $L_\mathrm{u}$ or a lower arm with length $L_\mathrm{d}$. 
After the reflection or transmission at QPC R they contribute to the current at contacts $2$ or $3$. 
The MZI is penetrated by a magnetic flux $\Phi$. 
As discussed in detail in ref.~\cite{Haack11}, coincidence at an MZI output of wave packets which traveled along different paths leads to an interference pattern in the current in contact $2$ or $3$. 
Particles from source B are emitted directly into the lower arm of the interferometer and can therefore not lead to any magnetic flux dependence in the current. 
Without  restriction of generality, we place cavity B at the center of the lower arm.

The single-particle sources~\cite{Feve07,Moskalets08} A and B are  small confined regions with discrete spectra weakly coupled to the conductor by QPCs, indicated by filled triangles in fig.~\ref{fig_setup}.
Uniform potentials are applied by top gates and modulated periodically in time with the same period $\mathcal{T} = 2\pi/\Omega$ for both sources.
We choose the amplitude of the modulations to be large, such that one level is driven back and forth through the Fermi level of the leads $\mu$. 
Then, well-separated electron and hole pulses are emitted from the source during each period~\cite{Moskalets08}. 
The parameters of each potential can be varied to tune the difference of times when particles are emitted.
 
We are interested in the regime of adiabatic driving, meaning that the time a particle spends in the cavity is small compared to the time, during which the quantum level of the source crosses the Fermi level~\cite{Splett08}. This results in the emission of a Lorentzian-shaped  current pulse~\cite{Olkhovskaya08}.
For the calculation of the adiabatic current emitted by the source $\alpha = \mathrm{A, B}$ at zero temperature, $I_\alpha(t) =  - ie/(2\pi) S_\alpha(t) \partial S^*_\alpha(t)/ \partial t$, \cite{Moskalets02} we need the instantaneous scattering amplitude calculated at the Fermi energy $\mu$ in the vicinity of the emission times ($0 < t \leq {\cal T}$)~\cite{Olkhovskaya08},
\begin{eqnarray}\label{eq_smatrix_ad}
S_\alpha(t) & = & \frac{\pm(t-t^\pm_\alpha)-i\Gamma_\alpha}{\pm(t-t^\pm_\alpha)+i\Gamma_\alpha} \,.
\end{eqnarray}
%\ \\
\noindent
Here $\Gamma_\alpha$ is the half width  and $t_\alpha^\pm$ is the emission time of an electron ($-$) and a hole ($+$) pulse. 
Their explicit dependence on the parameters of the source $\alpha = \mathrm{A, B}$ and its driving potential are given in ref.~\cite{Splett08}. 
Alternatively, single electrons can be emitted by means of Lorentzian voltage pulses~\cite{Keeling06} applied to the reservoir $1$. This is described via a similar scattering amplitude. 

While the modulation of the source alone is adiabatic, the time scales for the traversal of the MZI can be of the order of the period of the modulation $\mathcal{T}$. The time a particle needs for the traversal of the upper arm, $\tau_\mathrm{u}$, and the lower arm, $\tau_\mathrm{d}$, is directly related to the arm  lengths, $\tau_\ell=L_\ell/v_\mathrm{D}$ for $\ell=\mathrm{u,d}$, where $v_\mathrm{D}$ is the drift velocity of electrons along the edge state~\cite{McClure09,Chung05}. 
We relate the interferometer imbalance to the difference of traversal times $\delta\tau=\tau_\mathrm{u}-\tau_\mathrm{d}$.

%%%%%%%%%%%%%%%%%%%%%%%%%%%%%%%%%%%%%%%%%%%%%%%%%%%
\section{Results}
To find the current through the MZI, say at contact $2$, the full scattering matrix of the interferometer setup including the driven single-particle sources has to be constructed~\cite{Splett09}. 
With eq.~(\ref{eq_smatrix_ad}) this  leads to
\begin{subequations}
\begin{eqnarray}
I_{2}(t) & = & I_{2}^{(0)}(t)+I_2^{(\mathrm{int})}(t)\ ,\label{eq_current_tot}
\end{eqnarray}
%\ \\
consisting of a classical contribution and an interference part, which depends on the magnetic flux. The classical contribution is a sum of current pulses, 
\begin{eqnarray}
I_{2}^{(0)}(t) & = & I_\mathrm{A}(t-\tau_\mathrm{u})R_\mathrm{L}R_\mathrm{R}+
I_\mathrm{A}(t-\tau_\mathrm{d})T_\mathrm{L}T_\mathrm{R}\nonumber\\
\nonumber\\
&&+
I_\mathrm{B}(t-\frac{\tau_\mathrm{d}}{2})T_\mathrm{R}\ .\label{eq_current_classic}
\end{eqnarray}
%\ \\
Pulses emitted from source A arrive in a time interval around  $t^\pm_\mathrm{A}+\tau_\mathrm{u}$, respectvely $t^\pm_\mathrm{A}+\tau_\mathrm{d}$,  when traveling along the upper (respectively the lower) arm and pulses emitted from source B arrive in an interval around $t^\pm_\mathrm{B}+\tau_\mathrm{d}/2$. 
The interference part is given by
\begin{eqnarray}
I^{(\mathrm{int})}_2(t) & = & \frac{e\gamma}{\pi\left( \tau_\mathrm{d}-\tau_\mathrm{u} \right)}\mathrm{Im} \Big\{
S_\mathrm{B}^*\left(t-\frac{\tau_\mathrm{d}}{2}\right) \mathrm{e}^{i\phi} \nonumber \\
\nonumber\\
&&\times \left[S_\mathrm{A}(t-\tau_\mathrm{u})S^*_\mathrm{A}(t-\tau_\mathrm{d})-1\right] \Big\}\ ,
\label{eq_current_int}
\end{eqnarray}
\end{subequations}
%\ \\
where we defined $\gamma=\sqrt{R_\mathrm{L}R_\mathrm{R}T_\mathrm{L}T_\mathrm{R}}$ with $R_{\beta}(T_{\beta})$ being the reflection(transmission) coefficient of the QPC $\beta$, with  $\beta=\mathrm{L},\mathrm{R}$. Furthermore the phase is $\phi = 2\pi \Phi/\Phi_{0} + k_{\mu}v_\mathrm{D} \delta\tau$, with the magnetic flux quantum $\Phi_{0} = h/e$ and  $k_{\mu}$ is the wave number at the Fermi energy $\mu$.
We consider the regime where current pulses of electrons and holes from the same source are well separated, $\left| t_\alpha^+-t_\alpha^-\right| \gg\Gamma_\alpha$.

%%%%%%%%%%%%%%%%%%%%%%%%%%%%%%%%%%%%%%%%%%%%%%%%%%%
Let us first switch off the emitter B, resulting in  $S_\mathrm{B}(t)\equiv 1$ in eq.~(\ref{eq_current_int}).  
If the interferometer imbalance $\delta \tau$ is much larger than the pulse width $\Gamma_\mathrm{A}$, defining a first order coherence length~\cite{Haack11,Mandel99}, then the interference term is suppressed and the classical part, the first line of eq.~(\ref{eq_current_classic}), is the only contribution. 
We are now interested in the case $\delta\tau\ll \Gamma_\mathrm{A}$ such that interference \textit{is} present. 
We study the effect of  emitter B on the interference pattern of the signal from emitter A passing through the interferometer. 

%%%%%%%%%%%%%%%%%%%%%%%%%%%%%%%%%%%%%%%%%%%%%%%%%%%
\begin{figure}[b!]
\includegraphics[width=3.3in]{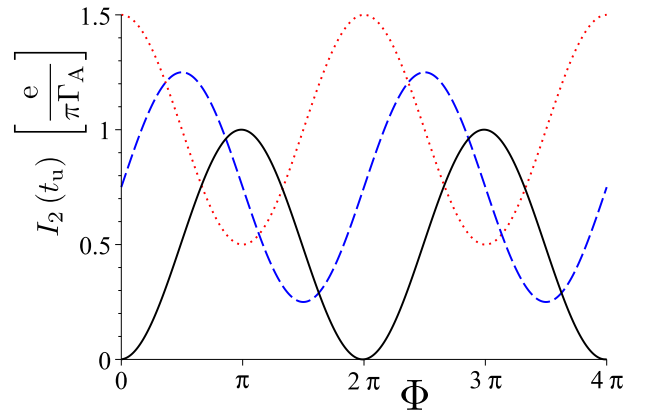}
\caption{ 
(color online). Current $I_{2}$, eq.~(\ref{eq_current_tot}), at a fixed time $t_\mathrm{u}=t_\mathrm{A}^-+\tau_\mathrm{u}$ as a function of the magnetic flux $\Phi$.  Results are shown for $\Delta t_\mathrm{u}=0$ (red, dotted line), $\Delta t_\mathrm{u}=\Gamma_\mathrm{A}$ (blue, dashed line), $\Delta t_\mathrm{u}\gg\Gamma_\mathrm{A}$ (black, full line). Furthermore we set $\delta\tau=0$ and $\Gamma_\mathrm{A}=\Gamma_\mathrm{B}$; QPCs are symmetric $R_\mathrm{L}=R_\mathrm{R}=0.5$.
\label{fig_current}
} 
\end{figure}
%%%%%%%%%%%%%%%%%%%%%%%%%%%%%%%%%%%%%%%%%%%%%%%%%%%

%%%%%%%%%%%%%%%%%%%%%%%%%%%%%%%%%%%%%%%%%%%%%%%%%%%
\section{Effect of source B on the current} 
Switching on emitter B, we obtain an additional contribution to the classical part of the current close to $t=t^\pm_\mathrm{B}+\tau_\mathrm{d}/2$, see eq.~(\ref{eq_current_classic}).  Concerning the interference part of the current, we see from eqs.~(\ref{eq_smatrix_ad}) and (\ref{eq_current_int}) that the effect of emitter B  is the addition of a phase. Importantly, this phase depends on time in the vicinity of a particle emission from B.
This is shown in fig.~\ref{fig_current}, where we plot the total current at a fixed measuring time $t_\mathrm{u}:=t^-_\mathrm{A}+\tau_\mathrm{u}$ as a function of the magnetic flux $\Phi$ for $\delta\tau=0$ for different values of $\Delta t_\mathrm{u}:=t^-_\mathrm{A} +\tau_\mathrm{u} - (t^-_\mathrm{B}+\tau_\mathrm{d}/2)$. 
For $\Delta t_\mathrm{u}\gg\Gamma_\mathrm{A}$, the signal from emitter B is negligible at time $t_\mathrm{u}$ and we find a current solely due to the source A, oscillating between $e/\left(\pi\Gamma_\mathrm{A}\right)$ and $0$ as a function of the magnetic flux. 
Decreasing the time-difference $\Delta t_\mathrm{u}$ leads to an increase of the average value of $I_{2}(t_\mathrm{u})$ due to the current stemming from B. 
A  phase shift, depending strongly on the value of $\Delta t_\mathrm{u}$, is introduced, without influencing the amplitude of the oscillations. This phase shift can take all values between $0$ and $2\pi$.

%%%%%%%%%%%%%%%%%%%%%%%%%%%%%%%%%%%%%%%%%%%%%%%%%%%
\section{Effect of source B on the charge} 
We consider here the regime, where each source emits one electron during the first half cycle and one hole during the second, i.e., integrating $I_{\alpha}(t)$ over the first half cycle, we obtain the charge $Q_\alpha= e$. 
We now want to study the charge $Q_2$ which is detected at contact $2$ by integrating the current $I_{2}(t)$ over the half cycle, in which electron pulses, emitted from both A and B, may be detected at contact $2$. Importantly, we integrate over a time interval much larger than the width of the current pulses, $\Gamma_{\alpha}$. The measured charge is 
\begin{subequations}
\begin{eqnarray}
Q_{2} & = & \int_{t_\mathrm{u}-\mathcal{T}/4}^{t_\mathrm{u}+\mathcal{T}/4}dt I_2(t)= Q^{(0)}_2+Q^\mathrm{(int)}_2  \label{3a}
\end{eqnarray}
%\ \\
with the classical $(0)$ and the interference (int) part
\begin{eqnarray}
Q^{(0)}_2 & = &  e(R_\mathrm{L}R_\mathrm{R}+T_\mathrm{L}T_\mathrm{R} + T_\mathrm{R})\,, \label{eq_averageQ} \\
\nonumber\\
Q^\mathrm{(int)}_2  & = &  
2e\gamma 
L(\Gamma_\mathrm{A};\delta\tau)
               \left[a\cos\phi+b\sin\phi\right]\,.
\label{eq_Q_ee}
\end{eqnarray}
\end{subequations}
%\ \\
In this equation, we defined the Lorentzian function $L(\Gamma;X)=4\Gamma^{2}/(4\Gamma^2+X^2)$ and the coefficients $a$ and $b$ are  given by
\begin{subequations}
\begin{eqnarray*}
a & = & -1+ \frac{\Gamma_\mathrm{B} }{\Gamma }
      L\left(
             \Gamma;\Delta t_\mathrm{u}
                    \right)
 \left[
         1-\frac{ \Delta t_\mathrm{u}
                     \delta\tau                 }{
                  4\Gamma_\mathrm{A}\Gamma
                 }
           \right]\label{eq_Q_eeA}\,, \\
           \nonumber \\
b & = & 
\frac{
     \delta\tau
    }{
       2\Gamma_\mathrm{A}
     }
\left[
       1-\frac{\Gamma_\mathrm{B} }{\Gamma } L\left(\Gamma;
                                                                 \Delta t_\mathrm{u}
                                                                \right)
\right]
-\frac{\Gamma_\mathrm{B}\Delta t_\mathrm{u} }{2\Gamma^{2} }
L\left(\Gamma;\Delta t_\mathrm{u} \right) .
\nonumber\\
\label{eq_Q_eeB}
\end{eqnarray*}
\end{subequations}
%\ \\
The quantity $\Gamma=(\Gamma_\mathrm{A}+\Gamma_\mathrm{B})/2$ is a mean half width.
To characterize how the detected charge $Q_2$ oscillates in magnetic flux $\Phi$, we introduce a  visibility $\nu=\left(Q_2^\mathrm{max}-Q_2^\mathrm{min}\right)/\left(Q_2^\mathrm{max}+Q_2^\mathrm{min}\right)$, the ratio of the amplitude of oscillations to the mean value around which the oscillation takes place. 

In general, if electrons emitted from A and B do \textit{not} meet each other at QPC R, the visibility is given by
\begin{eqnarray}
\label{eq_visibility} 
\nu_{0} &=& \frac{Q^{(0,\mathrm{A})}_2 }{Q^{(0)}_2 }\, \nu_\mathrm{A}(\delta\tau) \, .
\end{eqnarray}
Here $\nu_\mathrm{A} (\delta\tau) = 2e\gamma \sqrt{L(\Gamma_\mathrm{A}; \delta \tau) } / Q^{(0,\mathrm{A})}_2 $ is the visibility obtained for source A only, depending on the interferometer imbalance. 
The first factor, with $Q^{(0,\mathrm{A})}_2 = e(R_\mathrm{L}R_\mathrm{R}+T_\mathrm{L}T_\mathrm{R})$, is simply due to an increase of the classical charge contribution, when B is switched on and hence the average charge, $Q^{(0)}_2$, see eq. (\ref{eq_averageQ}),  is increased.

%%%%%%%%%%%%%%%%%%%%%%%%%%%%%%%%%%%%%%%%%%%%%%%%%%%
\section{Electron-electron collisions} 
However, a collision at QPC R of  an electron emitted by A, which traversed the upper arm of the MZI, with an electron emitted by B, suppresses the oscillations, even when the interferometer imbalance is zero. Such a collision leads to a decrease of the visibility, $\nu = \nu_{0} D_\mathrm{B}(\Delta t_\mathrm{u})$,  by a damping factor 
\begin{equation}
\label{dumping} 
D_\mathrm{B} = \sqrt{ \left( 1 - \frac{\Gamma_\mathrm{B} }{\Gamma } L(\Gamma; \Delta t_\mathrm{u}) \right)^{2} + \left( \frac{\Gamma_\mathrm{B} \Delta t_\mathrm{u} }{2\Gamma^{2} } L(\Gamma; \Delta t_\mathrm{u})\right)^{2}} \, ,
\end{equation}
%\ \\
which reduces the oscillation amplitude with increasing overlap,  $\Delta t_\mathrm{u} \ll \Gamma$. This time difference,  $\Delta t_\mathrm{u}$, characterizes the degree of overlap  at the QPC R between a current pulse from A traveling along the upper arm and a current pulse emitted from B.
If the current pulses overlap completely, i.e. the collision is perfect, $\Gamma_\mathrm{A} = \Gamma_\mathrm{B}$ and $\Delta t_\mathrm{u} = 0$, the oscillations vanish, since $D_\mathrm{B} =0$, and hence the visibility is zero. 
The factor $L^{1/2}(\Gamma_\mathrm{A}; \delta \tau)$ in $\nu_\mathrm{A}$ describes the suppression of interference at $\delta \tau \gg \Gamma_\mathrm{A}$ due to a decreasing overlap of wave packets injected from A traversing different interferometer arms~\cite{Haack11}.
These two effects causing an interference suppression enter the visibility as multiplicative factors and are therefore statistically independent.  
We emphasize, that the increase of the interferometer imbalance, $\delta \tau \gg \Gamma_\mathrm{A}$ would also suppress oscillations of the time-dependent current $I_{2}(t_\mathrm{u})$, discussed before for $\delta\tau=0$.  
However, the important result is that the collisions, suppressing $Q^\mathrm{(int)}_2 $, do not suppress the \textit{current oscillations} at all.

%%%%%%%%%%%%%%%%%%%%%%%%%%%%%%%%%%%%%%%%%%%%%%%%%%%
\begin{figure}[b!]
\includegraphics[width=3.3in]{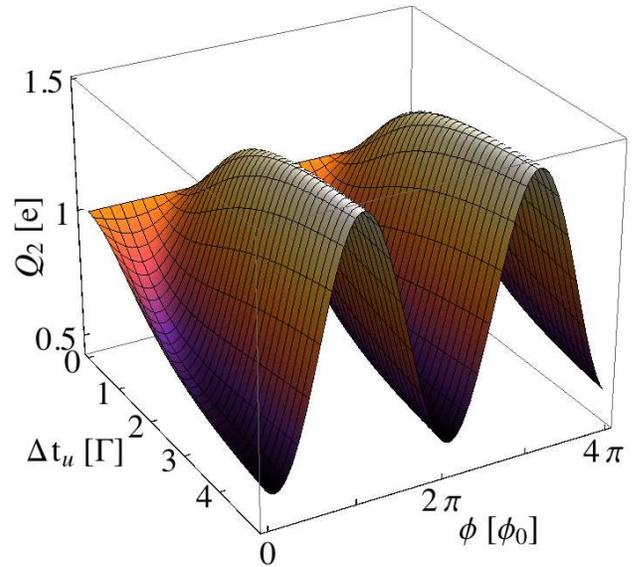}
\caption{ 
(color online). 
Detected charge $Q_2$, eq.~(\ref{3a}), as a function of the time difference $\Delta t_\mathrm{u}$ and the magnetic flux $\Phi$.
The interference contribution is fully suppressed if the collision condition $\Delta t_\mathrm{u}=0$ is fulfilled. 
Other parameters are the same as in fig.~\ref{fig_current}.
} 
\label{fig_charge}
\end{figure}
%%%%%%%%%%%%%%%%%%%%%%%%%%%%%%%%%%%%%%%%%%%%%%%%%%%

The detected charge $Q_2$ as a function of both the magnetic flux $\Phi$ and the time difference $\Delta t_\mathrm{u}$ is shown in fig.~\ref{fig_charge}. 
The AB oscillations are suppressed if the collision condition, $\Delta t_\mathrm{u} = 0$, is satisfied. 
The mechanism destroying the Aharonov-Bohm effect in the detected charge, $Q_2$, can be understood based on the particle-nature of the electron and it is related to which-path detection~\cite{Buks99,Dressel11}.  
The information on the path taken by the particle emitted from source A can be extracted from the two-particle state arriving at the detectors. 
Depending on the path taken by an electron incident from source A, two different outgoing two-particle states are possible. 
They differ from each other by whether the number of particles arriving at the detectors fluctuates or not.  
Consider an electron, emitted by the source A, which  is reflected at the QPC L  and which  traverses the interferometer along the upper arm. This occurs with probability $R_\mathrm{L}$.  
Under the condition $\Delta t_\mathrm{u} = 0$  it collides with an electron, emitted by the source B, at QPC R. 
Due to the Pauli principle, these two electrons can not be scattered into the same interferometer output. 
Therefore, the two particles arrive necessarily at different detectors~\cite{Olkhovskaya08} and
no fluctuations in the number of particles at the detectors are present in this state. The information on the origin of the two particles is lost  in this scattering process.
Otherwise, if an electron from A takes the lower arm, happening with probability $T_\mathrm{L}$, then both electrons pass the QPC R independently.
In this case the number of electrons arriving at each detector fluctuates. 
Note that, if $\Delta t_\mathrm{u} \gg \Gamma$,  then electrons from A and B are scattered at the QPC R independently of the arm through which the electron from A propagates. 
As a consequence the number of particles arriving at the detector fluctuates in both cases, meaning that no which-path information is acquired  and, therefore, the single-particle interference is not suppressed.

%%%%%%%%%%%%%%%%%%%%%%%%%%%%%%%%%%%%%%%%%%%%%%%%%%%
\section{Electron-hole annihilations}
Finally, we consider the case where the gate voltages at the sources are modulated such that an electron from source A arrives at the interferometer outputs in the same time interval as a hole emitted from source B. \par
Again, the time-resolved current experiences a time-dependent phase shift in the AB oscillations, while its average value is here lowered to $0$ due to an opposite sign in the contribution caused by the hole. We stress that also in this regime, no suppression of the interference in the \textit{current} occurs.\par
When integrating over the half cycle, in which electron and hole pass the QPC R, we find for the visibility of the charge the same damping factor as given by  eq.~(\ref{dumping}), with $\Delta t_\mathrm{u}$ substituted by $\Delta t_\mathrm{d} = t_\mathrm{A}^-+\tau_\mathrm{d} - ( t_\mathrm{B}^+ + \tau_\mathrm{d}/2)$. 
The interference part of the detected charge is therefore maximally suppressed if $\Delta t_\mathrm{d} = 0$, which is again due to which-path detection: If the electron from source A travels along the lower arm, it annihilates with a hole pulse from source B and no signal can be detected. 
In contrast, if an electron from source A travels along the upper arm, then this electron and the  hole emitted by the source B are scattered at QPC R independently. 
Therefore, in the latter case the electrons and holes will arrive at the detector and their number will fluctuate.  This difference in the outgoing two-particle states leads to a suppression of the magnetic-flux dependence.

\section{Conclusion}
We calculated the time-resolved current and the charge per half cycle at the output of a Mach-Zehnder interferometer due to injection from single-particle sources. We showed that these quantities differ fundamentally with respect to their sensitivity to the magnetic flux.  This allows for an interpretation relying either on the wave-like or on the particle-like nature of the emitted electrons within the same setup. The current oscillates in magnetic flux witnessing coherence. In contrast, the oscillations in the transferred charge can be suppressed via electron collisions or electron-hole annihilations. This is the signature of a two-particle effect becoming visible in a single-particle quantity, which is an important finding for the coherent control of few-electron quantum states.

\acknowledgments 
We  thank  M. B\"uttiker, P. Samuelsson, E. Sukhorukov, and U. Z\"ulicke for useful discussion.
We acknowledge financial support from the Ministry of Innovation NRW.

%%%%%%%%%%%%%%%%%%%%%%%%%%%%%%%%%%%%%%%%%%%%%%%%%%%

\end{document}